\documentclass[aps,prb,showpacs,twocolumn,preprintnumbers,amsmath,amssymb,superscriptaddress]{revtex4}

\usepackage{graphicx}
\usepackage{dcolumn}
\usepackage{bm}
\begin{document}

\title{Bounds for the Superfluid Fraction from Exact Quantum Monte Carlo Local Densities}
\author{D.E. Galli}
\affiliation{ Dipartimento di Fisica, Universit\`a degli Studi di Milano, Via Celoria 16, 20133 Milano, Italy}
\author{L. Reatto}
\affiliation{ Dipartimento di Fisica, Universit\`a degli Studi di Milano, Via Celoria 16, 20133 Milano, Italy}
\author{W.M. Saslow}
\affiliation{ Department of Physics, Texas A\&M University, College Station, TX 77843-4242}
\date{\today}

\begin{abstract}
For solid $^4$He and solid p-H$_2$, using the flow-energy-minimizing one-body phase function and exact 
$T=0$ K Monte Carlo 
calculations of the local density $\rho({\vec r})$, we have calculated the phase function, the velocity profile
and upper bounds for the superfluid fraction $f_s$.
At the melting pressure for solid $^4$He we find that $f_s \leq 0.20-0.21$, about ten times what is observed.
This strongly indicates that the theory for the calculation of these upper bounds needs substantial improvements.
\end{abstract}


\maketitle

The recent observation of non-classical rotational inertia (NCRI) effects
in bulk solid $^4$He,\cite{KCbulk}
and the experimental investigation of condensates
in optical lattices\cite{lattice} have increased theoretical interest in the study of superfluid
properties in presence of translational broken symmetry.
The interpretation of NCRI effects is still controversial and
arguments against bulk supersolidity in solid $^4$He have appeared in the literature.\cite{cepe,proko}
To date there are microscopic indications of supersolidity from exact simulations only 
in presence of a finite concentration of vacancies.\cite{Gallinew}
The presence of ground state vacancies in bulk solid $^4$He 
also is controversial.\cite{Gallinew,Boninew}
Superfluidity of interfaces have been also invoked\cite{proko},
but the quantitatively similar presence of NCRI effects in solid $^4$He confined in different porous
media\cite{KCvycor,KCgold} seems to be difficult to reconcile with this single mechanism.

Here we return to the $T=0$ K analysis initiated by Leggett in which an upper bound
for the superfluid fraction $f_s$ was deduced.
In 1970, Leggett studied superflow in a solid confined to an annulus.\cite{Leggett1}
He considered that, under rotation, 
the ground state wave function $\Psi_{0}$ became $\Psi=\Psi_{0}\exp({i\phi})$, 
with a phase function $\phi$ that was a sum of terms depending only upon a single
longitudinal coordinate (what we call a one dimensional one-body phase function).
He noted that this would yield only an upper bound to the
$T=0$ K superfluid fraction $f_{s}$, but he did not evaluate it.
This bound depends on
the averaged density, $\rho(u)=\int d\xi \rho({\vec r})$, where
$\rho({\vec r})$ is the local density,
$u$ is the longitudinal coordinate and $\xi$ is a suitable set of transversal coordinates.
By modelling $\rho({\vec r})$ 
as a sum of Gaussians centered on the lattice sites -- 
the so-called Gaussian Model (GM) -- Ref.~\onlinecite{puma} evaluated Leggett's bound. 
Ref.~\onlinecite{Saslow1} extended Leggett's bound by using the GM and an improved variational ansatz: 
the phase is a function of ${\vec r}$, not merely of the
longitudinal coordinate.  
The most recent calculations\cite{Saslow2,Saslow-hcp}
gave an upper bound for $f_s$ on the order of 2\%, in good agreement with experiments on solid $^4$He. 
This bound, however, depends very strongly on the value of the width of the Gaussians
so that this result is model dependent. 

The present work reports calculations of upper bounds for $f_{s}$
that, for the first time, use consistent $T=0$ K densities computed 
with modern exact Quantum Monte Carlo (QMC)
techniques.
The new results are very different:
the exact QMC one-body density gives an upper bound on the order of 20\%,
for $^4$He near the melting density.
This is some ten times higher than
observed experimentally.  Independent of the interpretation of NCRI effects
in solid $^4$He, this indicates that the theory for calculatiing these upper bounds
should be re-examined, at least for strongly interacting systems.
This judgment could be more negative in case the NCRI effects measured in solid $^4$He are not
an equilibrium property but are induced
only by the presence of disorder in the system.

Ref.~\onlinecite{Saslow1} showed that a one--body phase function $\phi(\vec{r})$ minimizes
the flow energy $E=m/2 \int d\vec{r} \rho(\vec{r}) v_s^2(\vec{r})$,
where $v_s(\vec{r})=\hbar/m\vec{\nabla}\phi(\vec{r})$,
when the continuity equation
\begin{equation}\label{ceq}
\vec{\nabla}\cdot[\rho(\vec{r}) v_s(\vec{r})]=0
\end{equation}
is satisfied. Thus, on imposing a uniform velocity $v_0$ on a known $\rho(\vec{r})$, one can obtain $v_s(\vec{r})$ and then an upper bound for the superfluid density from
\begin{equation}\label{uppb} 
\rho_s v_0^2 V \leq \int d\vec{r} \rho(\vec{r}) v_s^2(\vec{r}) \quad .
\end{equation}
We have computed the local density $\rho(\vec{r})$ in solid $^4$He and in 
solid p-H$_2$ using an extension of the Path Integral Ground State method.\cite{sch}
Specifically, we compute 
$\rho(\vec{r}) = \langle \Psi_0 \mid \sum_{i=1}^N \delta(\vec{r}-\vec{r}_i) \mid \Psi_0 \rangle$,
using the exact $T=0$ K projector Shadow Path Integral Ground State (SPIGS) method.\cite{gal1}
In a projector method the exact ground state
is expressed as the imaginary
time evolution of a trial variational state; in the SPIGS method this trial state is chosen
to be a Shadow Wave Function (SWF).\cite{swf1,swf2}
It is well documented\cite{swf3} that a SWF gives presently the best variational representation
of $^4$He both in the liquid and in the solid phase.
With $R\equiv\{\vec{r}_1,..,\vec{r}_N\}$ representing the many-body coordinates, we have\cite{swf1}:
\begin{equation}
\Psi_0(R) = \lim_{\tau \to \infty} \int dR' \, G(R,R',\tau) \Psi^{SWF}(R') \quad,
\end{equation}
where $G(R,R',\tau)=<R|\exp(-\tau \hat{H})|R'>$
is the exact imaginary time propagator and $\Psi^{SWF}(R')$ is the
SWF.
                                                                                                                   
In general, the exact $G(R,R',\tau)$ is not known;
however, via the path integral representation one can express
$G(R,R',\tau)$ as a convolution of many (for example, $M$)
short imaginary time propagators $G(R,R',\delta\tau)$,
where $\delta\tau=\tau/M$, and accurate
approximations are available for $G(R,R',\delta\tau)$.
The imaginary time evolution is stopped when $\tau$ is sufficently large that averages
are stable with respect to the value of $\tau$.  This means that $\Psi(R,\tau)$
has essentially zero overlap with the excited states.
In this way averages computed with $\Psi^{2}(R,\tau)$ for the quantum system of $N$ 
interacting atoms are
equivalent to canonical averages for a classical system of
special interacting open polymers of $2M+1$ particles per actual atom,
where $M$ is the number of projection steps
(convolutions) in imaginary time.\cite{gal1}
Ground state averages of diagonal operators
can be computed via an efficient Monte Carlo sampling of the many--body coordinates
for the imaginary time path; this is obtained
by interpreting the short imaginary time propagators and the trial state
as probability densities in an extended space of $N(2M+1)$ coordinates.
                                                                                                                   
In the SPIGS method the imaginary time evolution starts from a SWF, which can  
describe both the liquid
and the solid phase with the same functional form:
spatial order in the solid phase
is produced by spontaneously breaking the
translational symmetry of the Hamiltonian. 
(To compute the density
a special class of moves is made, discussed below).
 With $S\equiv\{\vec{s}_1,..,\vec{s}_N\}$ representing the set
of subsidiary ``shadow'' variables $\vec{s}_{i}$, we have:
$\Psi^{SWF}(R) = \int dS \, \, \phi(R) K(R,S) \phi_s(S)$.
Here $\phi(R)$ contains the explicit part of the interparticle correlations,
which are
assumed of the Jastrow form (two-body correlation function);
the kernel $K(R,S)$ is often taken as a product of gaussians
$K(R,S)=\prod_{i=1}^N e^{-C|{\vec r}_i-{\vec s}_i|^2}$,
but better representations are known,\cite{swf3}
that correlates each coordinate ${\vec r}_i$ with its associated shadow variable ${\vec s}_i$;
and $\phi_s(R)$ is another Jastrow factor, which accounts for correlations
between the shadow variables.
Integration over the
subsidiary (shadow) variables introduces, implicitly, effective correlations
between the particles, which are not limited to pair or triplet terms;
in principle, terms of all order are generated.
When the density of the system exceeds the melting density of the solid, such correlations
become so strong that the solid phase stabilizes with an energy below that
of the (now metastable) liquid phase\cite{swf1,swf2};
there is no need to introduce {\it ad hoc}
equilibrium positions to locate the atoms in the solid phase, as usually done within
the standard variational theory.

The SWF technique has been shown to be very powerful but, as in all variational
computations, it is intrinsically limited by the
functional form of the
trial wave function. A way to overcome this
limitation is to use the SWF as a guiding function in an algorithm that
converges to the exact wave function.
This has been obtained with the SPIGS method, which inherits the above-mentioned properties
by projecting out the ground state wave function
from the imaginary time evolution of a SWF.\cite{gal1}
This is especially relevant in the study of a quantum solid, where exchange or more complex
zero-point phenomena could play a relevant role in the determination of properties 
like the local density\cite{GalliRR}.
Since our wave function is translationally invariant the local density turns out
to be a constant after a very large number of Monte Carlo steps due to the
fluctuations of the center of mass of the system. In order to mimic the behavior
of a macroscopic solid held fixed in the laboratory frame we perform the SPIGS 
computation at fixed center of mass: 
only moves that involve all the particles are proposed to the Metropolis acceptance test such that
when a particle ${\vec r}_i$ is moved by ${\vec \delta}$
all the others $N-1$ particles are moved by $-{\vec \delta}/(N-1)$. This is done for each time slice.

We have computed the local density $\rho(\vec{r})$ in hcp solid $^4$He near the melting density
($\rho=0.029$ \AA$^{-3}$), at $\rho=0.0353$ \AA$^{-3}$ and also in hcp solid p-H$_2$ at its
equilibrium density $\rho=0.026$ \AA$^{-3}$, where recent experimental results
seem to exclude the presence of NCRI effects.\cite{ph2}
The SWF used for the initial state of the imaginary time evolution is a standard
SWF\cite{swf1,swf2}.
Our SPIGS method employs the pair-product approximation\cite{pimc} for the imaginary time propagators. The imaginary time step used in these simulations was $\delta\tau=\frac{1}{80}$ K$^{-1}$. 
With $M=15$ time slices this gave an imaginary
projection time of $M\delta\tau=0.1875$~K$^{-1}$, and a total of 
33 particles in the open polymers  (this includes two shadow particles).
However, only the 
11 interior time-slices were used to calculate the ground state local density.
We have verified that such $\delta\tau$ is accurate for solid $^4$He
by checking the convergence of diagonal properties computed
also with $\delta\tau=\frac{1}{40}$, $\frac{1}{160}$ and $\frac{1}{320}$ K$^{-1}$.
The potential used to model the interaction between $^4$He atoms was a standard Aziz potential\cite{aziz},
whereas the potential of Silvera and Goldman\cite{silv} was used for p-H$_2$.

\begin{figure*}
\centerline {
\includegraphics[width=5.4in]{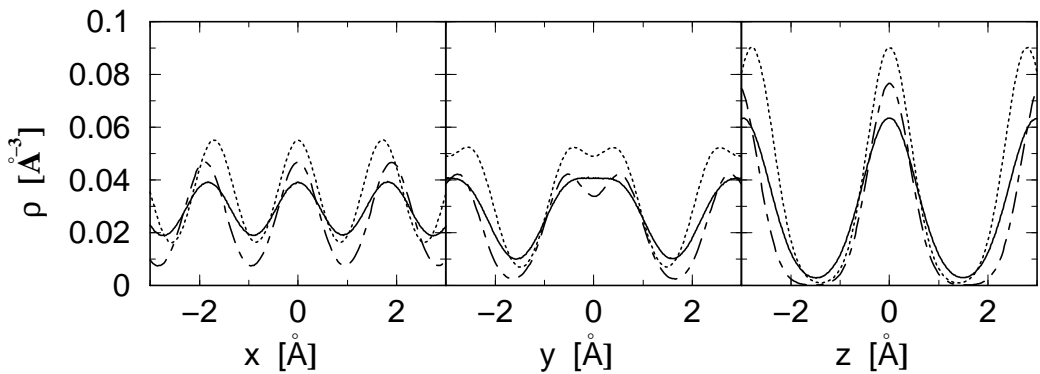}
\vspace{0in}}
\caption{Averaged local densities along the axis of the simulation 
box for hcp solid $^4$He at
$\rho=0.029$ \AA$^{-3}$ (solid line),  hcp solid $^4$He at $\rho=0.0353$ \AA$^{-3}$ (dotted line),
hcp solid p-H$_2$ (dot-dashed line) at $\rho=0.026$ \AA$^{-3}$. The basal plane is perpendicular to
the z direction; the x axis and y axis are respectively parallel to the $\Gamma$K and $\Gamma$M direction.
The choice of the origin along the axis is arbitrary. The periodicity of the oscillations in
the averaged local densities depends on the lattice constant and changes with the average density $\rho$.}
\label{grafico}
\end{figure*}

We first use our QMC densities to evaluate the less accurate upper bound for $f_{s}$ formulated in 
Ref.~\onlinecite{Leggett1}, based on a one-body phase function that depends on one spatial dimension.  
The averaged local one-dimensional densities ${\rho}(u)$
along the three axes $u=x,y,z$ of the hcp simulation 
box are shown in Fig.\ref{grafico};
they correspond to the local density $\rho(\vec{r})$ integrated along the plane perpendicular to
the chosen axis.
$\rho(\vec{r})$ has been computed on $64^3$ real-space points in a cell 
of dimension $(a,\sqrt{3}a,\sqrt{8/3}a)$. There are 45 equivalent cells of this
kind in the hcp simulation box with N=180 atoms; averaging the
local densities in each cell has been used in order to improve the statistics.
Fig.\ref{grafico} shows that, of the three ${\rho}(u)$'s, ${\rho}(z)$ (for the direction perpendicular 
to the basal plane) has the strongest oscillations around the average density $\rho$ in the system.
This method of averaging includes peaks from different planes;
for $\rho(y)$, the peaks from different planes overlap.
As seen in Fig.\ref{grafico}, near melting ($\rho=0.029$ \AA$^{-3}$) the peaks are so broad that
the average gives a single broadened ``peak'' in $\rho(y)$.
However, at higher pressure, where the atoms are more localized
($\rho=0.0353$ \AA$^{-3}$), the contributions of the individual planes can be partially resolved.
For each of the three cases depicted in Fig.\ref{grafico}, we have computed the upper bounds 
for flow along $x$, $y$, and $z$.  
In each case, the lowest of these upper bounds is associated with flow along $z$, 
because the large oscillations in $\rho(z)$ give deeper minima.
These lowest 1d upper bounds are  $f_s \leq 0.384$ for hcp solid $^4$He at
$\rho=0.029$ \AA$^{-3}$ (flow along $x$ gives $f_{s}\le 0.939$ and flow along $y$ gives $f_{s}\le0.799$), 
$f_s\leq 0.164$ for hcp solid $^4$He at
$\rho=0.0353$ \AA$^{-3}$ (flow along $x$ gives $f_{s}\le 0.839$ and flow along $y$ gives $f_{s}\le0.648$) 
and $f_s \leq 0.054$ for hcp solid p-H$_2$ 
at $\rho=0.026$ \AA$^{-3}$ (flow along $x$ gives $f_{s}\le 0.686$ and flow along $y$ gives $f_{s}\le0.458$). 

An improved upper bound for $f_{s}$ is obtained by using a one-body phase
function that depends on all three spatial dimensions and the
local density $\rho(\vec{r})$ given by SPIGS.
Note that the one-body phase is the sum of an 
imposed uniform flow and a one-body backflow that varies on the atomic scale 
as follows from Eq.(\ref{ceq}).
Eqs.(\ref{ceq}-\ref{uppb}) are solved in Fourier space.
We have studied $f_s$ for the hcp lattice, as actually realized experimentally.  In this case the 
crystal structure is a Bravais lattice with a basis of two.  By placing the origin midway 
between the basis atoms, one may employ only real fourier transforms.  
\begin{table*}
\caption{Superfluid fraction upper bound, computed with SPIGS and SWF, for hcp solid $^4$He
and hcp solid p-H$_2$ at different densities for varying basis sizes $(2P+1)^3$ in fourier space.}
\begin{center}\begin{tabular}{|c|cccc|cccc|cccc|}
\hline
  & \multicolumn{4}{l|}{\quad\quad\quad$^4$He,\quad $\rho=$0.029\AA$^{-3}$} &\multicolumn{4}{l|}{\quad\quad\quad$^4$He, \quad$\rho=$0.0353\AA$^{-3}$} &\multicolumn{4}{l|}{\quad\quad\quad p-H$_2$, \quad$\rho=$0.026\AA$^{-3}$}  \\
\cline{2-13}
$P$&$f_s$(100) &$f_s$(010) &$f_s$(001) &$f_s^{SWF}$(001)&$f_s$(100) &$f_s$(010) &$f_s$(001) &$f_s^{SWF}$(001)&$f_s$(100) &$f_s$(010) &$f_s$(001) &$f_s^{SWF}$(001)\\
\hline
1  &0.4055 &0.3801 &0.5694 &0.5015                      &0.2517 &0.2253 &0.3968 &0.3736          &0.1553 &0.1336 &0.2653 &0.2238          \\
2  &0.2884 &0.2733 &0.3340 &0.2590                      &0.1267 &0.1127 &0.1643 &0.1457          &0.0526 &0.0436 &0.0736 &0.0512          \\
3  &0.2464 &0.2389 &0.2618 &0.1852                      &0.0860 &0.0792 &0.1004 &0.0846          &0.0263 &0.0225 &0.0332 &0.0195          \\
4  &0.2310 &0.2267 &0.2346 &0.1573                      &0.0705 &0.0667 &0.0770 &0.0626          &0.0175 &0.0155 &0.0207 &0.0106          \\
5  &0.2249 &0.2219 &0.2224 &0.1444                      &0.0637 &0.0614 &0.0662 &0.0526          &0.0137 &0.0126 &0.0154 &0.0071          \\
6  &0.2224 &0.2199 &0.2166 &0.1379                      &0.0605 &0.0589 &0.0606 &0.0473          &0.0119 &0.0112 &0.0127 &0.0054          \\
7  &0.2213 &0.2190 &0.2137 &0.1345                      &0.0588 &0.0576 &0.0576 &0.0444          &0.0109 &0.0105 &0.0112 &0.0045          \\
8  &0.2208 &0.2186 &0.2124 &0.1327                      &0.0580 &0.0569 &0.0558 &0.0427          &0.0104 &0.0101 &0.0103 &0.0039          \\
\hline
\end{tabular}\end{center}
\label{hcp4He0.029}
\end{table*}
The results are given in Table~\ref{hcp4He0.029} for hcp $^4$He at $\rho=0.029$ \AA$^{-3}$,
for hcp $^4$He at $\rho=0.0353$ \AA$^{-3}$ and for hcp p-H$_2$ at $\rho=0.026$ \AA$^{-3}$;
the number of fourier components associated with each reciprocal lattice vector basis 
is given by $2P+1$.  
In the table the accuracy of the QMC density does not warrant four decimal places, but
the same QMC density was used for each calculation, so that relative values have four place accuracy.
The $f_s$ shown in Table~\ref{hcp4He0.029} have been obtained from $\rho(\vec{r})$ computed
with an imaginary projection time
of $M\delta\tau=0.1875$ K$^{-1}$; we have checked the convergence of these results
by computing the local density with
different imaginary projection times: $M\delta\tau=0.15$ K$^{-1}$ and $M\delta\tau=0.2625$ K$^{-1}$.
In Table~\ref{hcp4He0.029} we have shown also $f_{s}$ computed with SWF; the variational $f_{s}$
turn out to be always lower as a consequence of the larger degree of local order\cite{GalliRR}.
This is consistent with the larger Lindemann's
ratio $\gamma=0.257(4)$ obtained with SPIGS at $\rho=0.029$ \AA$^{-3}$
as opposed to the one, $\gamma=0.242(2)$, obtained with SWF;
the experimental value is $\gamma^{exp}=0.263(6)$.\cite{burns}
Note that, to within numerical accuracy, all
three systems are isotropic.
Such isotropy of $f_{s}$ need not hold for a hexagonal lattice, but near-isotropy is
consistent with a Gaussian Model (GM) for the hexagonal lattice.\cite{Saslow-hcp}
Although inspection of the tables indicates that these three systems are nearly isotropic,
in each case visual inspection of the data (connected by straight lines) indicates that $f_{s}(001)$,
associated with flow along $z$, is extrapolating to a value a few percent lower than for the other two.
Similarly, visual inspection indicates that $f_{s}(100)$ and $f_{s}(010)$ appear to extrapolate to
the same value,  indicating isotropicity for flow along the basal hcp plane.

Converged values for $f_s$ 
for $^4$He at melting density $\rho=0.029$ \AA$^{-3}$
($f_{s,xx}=0.220, f_{s,yy}=0.218, f_{s,zz}=0.211$) 
lie in the range $f_{s}=0.21-0.22$. 
In terms of the GM\cite{Saslow-hcp}, this corresponds to an rms width of
$\sigma \approx 0.1485 \, d$, where $d$ is the
nearest-neighbor distance for the hcp lattice.
Note that in the GM, $\sigma =0.1414 \, d$ gives $f_s=0.133$ and $\sigma =0.159 \, d$ gives 
$f_s=0.303$, so that $f_s$ is a very sensitive function of the localization.  
Variations in $f_s$ according to direction are likely
due to numerical uncertainties in the QMC density $\rho(\vec{r})$, of the order of
3\% at this density.
This uncertainty is unlikely to be the source of the factor of ten difference 
between the one-body calculation described above, and the experimental values on the order 
of 1-2\%.\cite{KCbulk}
The system is therefore less localized than previous calculations had indicated,\cite{Saslow2} and
this is reflected in the larger value of $f_{s}$, which increases with
the extent of delocalization, in principle to a maximum of unity (the fluid state).
A possible source of discrepancy lies also in the usage of the GM in the previous calculations:\cite{Saslow2}
this model in fact loses accuracy when we consider
the regions of the minima of $\rho({\vec r})$; here deviations
from the exact QMC local densities greater than 100\% are found.
However these discrepancies seem to have only a small effect on the
derived upper bounds $f_s$: by fitting the QMC averaged local densities with the GM we obtain
for $^4$He at $\rho=0.029$ \AA$^{-3}$ a rms width $\sigma = 0.1486 \, d$
and at $\rho=0.0353$ \AA$^{-3}$ a rms width $\sigma = 0.1274 \, d$; in the GM\cite{Saslow2,Saslow-hcp}
these rms widths correspond respectively to $f_s \approx 0.22$
and $f_s \approx 0.05$, in good agreement with the results in 
Table~\ref{hcp4He0.029}.
Solid $^4$He at $\rho=0.0353$ \AA$^{-3}$ and solid p-H$_2$ at $\rho=0.026$ \AA$^{-3}$ are more 
localized systems, and
give correspondingly lower values for $f_{s}$ 
(see Table~\ref{hcp4He0.029});
however even if there seems to be no indication of NCRI effects in solid p-H$_2$\cite{ph2} the upper bound
is still over 1\%.
We have also computed the upper bound for the superfluid fraction for $^4$He on an fcc lattice;
at 0.029 \AA$^{-3}$ the results are close to
those obtained for the hcp lattice.

To conclude, we have performed calculations of the upper bound for the superfluid fraction using an optimized 
one-body phase function and state-of-the-art
QMC one-body densities.  We find an upper bound for the superfluid fraction of 20\% 
for solid $^4$He at density of 0.029 \AA$^{-3}$ (near the melting pressure), significantly higher 
than what has been observed.
Therefore the usage of exact local densities has definitely shown that the accuracy of the upper bounds for the
superfluid fraction is of little utility in the present form, at least for a strongly interacting
quantum system like solid $^4$He and solid p-H$_2$; in the case of condensates in optical
lattice the conclusions could be different.
However, given an additional degree of freedom by permitting a many-body phase function, it 
should be possible to lower the flow energy for the case of the solid, and therefore to 
lower the corresponding superfluid fraction, perhaps by a considerable amount.  
We have developed such a theory\cite{condmat} for a two-body phase function, and the results should also be 
applicable to condensates in optical lattices and to disordered solids.

This work was supported by the INFM Parallel Computing
Initiative and by the Mathematics Department ``F. Enriques''
of the Universit\`a degli Studi di Milano, and by the Department of Energy through grant 
DE-FG02-06ER46278.  We would also like to acknowledge the hospitality of the 
Kavli Institute for Theoretical Physics and the Aspen Institute of Physics. 

{}

\end{document}